\documentclass[letterpaper, 10 pt, conference]{ieeeconf}  

\IEEEoverridecommandlockouts                              

\usepackage{graphics} 
\usepackage{epsfig} 
\usepackage{mathptmx} 
\usepackage{times} 
\usepackage{amsmath} 
\usepackage{amssymb}  
\usepackage{xcolor}
\usepackage{booktabs}
\usepackage{tabularx}
\usepackage{multirow}
\usepackage{arydshln}
\usepackage{siunitx}
\usepackage{cite}
\usepackage{hhline}

\usepackage[absolute]{textpos}
\usepackage{everyshi}

\title{\LARGE \bf
Deep Learning Techniques for Improving Digital Gait Segmentation
}

\author{ Matteo Gadaleta$^{1}$, Giulia Cisotto$^{1,2}$, Michele Rossi$^{1}$, Rana Zia Ur Rehman$^{3}$, Lynn Rochester$^{3,4}$, Silvia Del Din$^{3}$
\thanks{$^{1}$ Department of Information Engineering (DEI), University of Padova, Italy (e-mail: {\tt\small gadaleta,cisottog,rossi@dei.unipd.it}) }%
\thanks{$^{2}$ Integrative Brain Imaging Centre, National Centre of Neurology and Psychiatry (NCNP), Tokyo, Japan}%
\thanks{$^{3}$ Institute of Neuroscience/Newcastle University Institute for Ageing, Clinical Ageing Research Unit, Campus for Ageing and Vitality, Newcastle University, Newcastle upon Tyne, UK (e-mail: {\tt\small rana.zia-ur-rehman, silvia.del-din@ncl.ac.uk}) }%
\thanks{$^{4}$ Newcastle upon Tyne Hospitals NHS Foundation Trust (e-mail: {\tt\small lynn.rochester@newcastle.ac.uk}) }%
}

\newcommand{\fig}[1]{Fig.~\ref{#1}}

\begin{document}
\bstctlcite{BSTcontrol}

\maketitle
\thispagestyle{empty}
\pagestyle{empty}


\begin{abstract}
Wearable technology for the automatic detection of gait events has recently gained growing interest, enabling advanced analyses that were previously limited to specialist centres and equipment (e.g., instrumented walkway). 
In this study, we present a novel method based on dilated convolutions for an accurate detection of gait events (initial and final foot contacts) from wearable inertial sensors.
A rich dataset has been used to validate the method, featuring $71$ people with Parkinson's disease (PD) and $67$ healthy control subjects.
Multiple sensors have been considered, one located on the fifth lumbar vertebrae and two on the ankles.
The aims of this study were: (i) to apply deep learning (DL) techniques on wearable sensor data for gait  segmentation and quantification in older adults and in people with PD; (ii) to validate the proposed technique for measuring gait against traditional gold standard laboratory reference and a widely used algorithm based on wavelet transforms (WT); (iii) to assess the performance of DL methods in assessing high-level gait characteristics, with focus on stride, stance and swing related features. The results showed a high reliability of the proposed approach, which achieves temporal errors considerably smaller than WT, in particular for the detection of final contacts, with an inter-quartile range below \SI{70}{\ms} in the worst case.
This study showes encouraging results, and paves the road for further research, addressing the effectiveness and the generalization of data-driven learning systems for accurate event detection in challenging conditions.
\end{abstract}

\begin{textblock*}{17cm}(1.7cm, 0.5cm)
\noindent\scriptsize This paper will be published on \emph{2019 IEEE $41^{st}$ International Engineering in Medicine and Biology Conference (EMBC), Berlin, July 2019.}\\
\textbf{Copyright Notice}: \textcopyright 2019 IEEE. Personal use of this material is permitted. Permission from IEEE must be obtained for all other uses, in any current or future media, including reprinting/republishing this material for advertising or promotional purposes, creating new collective works, for resale or redistribution to servers or lists, or reuse of any copyrighted component of this work in other works.
\end{textblock*}

\section{INTRODUCTION}

Gait impairment is frequent among an aging population and in particular in neurodegenerative diseases, e.g., the Parkinson's disease (PD).  Gait performance is often considered as the sixth vital sign and it is emerging as a powerful tool to identify surrogate markers of incipient disease manifestation and disease progression~\cite{1_lord2013moving}.
Traditionally, gait analysis has been carried out using specialised equipment (most commonly instrumented walkways such as pressure-sensor activated, e.g., GAITRite)~\cite{6_rochester2014nature,8_williams2013gait}
which limits one to work within specialised centres and gathering a relatively small number of gait cycles~\cite{10_paterson2009gait}. Recent studies have shown that wearable technology (e.g., inertial sensors) is a valid and inexpensive alternative for quantifying digital gait outcomes both in controlled and daily living environments~\cite{11_godinho2016systematic,13_del2016free}.
Many approaches have been developed for gait segmentation (e.g., detection of initial and final contacts) and quantification of discrete gait characteristics, such as temporal, variability and asymmetry metrics. Amongst these, reliability and accuracy can vary with respect to protocol (e.g., sensor position, number of sensors) and method (peak detection, feature-based, template-based methods, etc.) as features depend on signal quality and, often, on the use of  thresholds~\cite{18_trojaniello2015comparative,19_del2016validation,20_pham2017validation}.
%
Recently, machine learning (ML) and deep learning (DL) techniques have been utilised for automatic gait segmentation and pathology classification to support clinical decision making, with good results~\cite{25_tahir2012parkinson,26_wahid2015classification,28_eskofier2016recent}.
Widely reported ML models in these studies are support vector machine (SVM), random forest, k-nearest neighbour, classification and regression trees, neural networks, and logistic regression. An advantage of DL over traditional ML techniques is their good generalization property, which allows for an increased reliability of signal segmentation, especially in unsupervised settings. Moreover, DL techniques successfully deal with high dimensionality and high variability data, which are frequent with wearable technology, and make the extraction of informative features in \mbox{real-time} possible (e.g., to help anticipate falls), enabling continuous monitoring systems.
Deep convolutional neural networks (CNN) have proven to be an effective method for the extraction of gait characteristics from inertial sensor data, taking advantage of hierarchical \mbox{non-linear} processing to learn \mbox{high-level} data representations (\textit{representation learning})~\cite{hannink2017sensor}. However, CNNs typically follow a preceding gait segmentation phase, which is thus critical to their effectiveness. This introduces the need for further investigations (and optimisations) on the segmentation methods for gait data from wearable sensors, before the clinical adoption of this technology.
%

Along these lines, the purpose of the present study is to apply CNNs and validate their use for the accurate segmentation of gait signals and the quantification of their salient features. In particular, we aim at: \textbf{(i)} applying DL techniques on wearable sensor data for gait segmentation in older adults and in people with PD; \textbf{(ii)} validating DL techniques for measuring gait against a widely used approach including acquisition via a gold standard laboratory reference (GAITRite), and data processing based on wavelet transform (WT); 
\textbf{(iii)} evaluating the performance of the proposed DL method in measuring high-level gait characteristics, with focus on stride, stance and swing related features.

The paper is structured as follows. The dataset and the proposed \mbox{CNN-based} algorithms are presented in Sec.~\ref{sec:methods}, quantitative results are shown in Sec.~\ref{sec:results}, and concluding remarks are given in Sec.~\ref{sec:conclusions}.

\section{METHODS}
\label{sec:methods}

\subsection{Data collection} 

A group of $71$ PD patients (age: $68.95\pm9.17$ years, MDS-UPDRS III: $40.63\pm11.90$) and $67$ healthy control subjects (HC, age $71.20\pm6.49$ years) were recruited from the Incidence of Cognitive Impairment in Cohorts with Longitudinal \mbox{Evaluation-GAIT} (ICICLE-GAIT) study. This is a collaborative study with ICICLE-PD, an incident cohort study (Incidence of Cognitive Impairment in Cohorts with Longitudinal Evaluation Parkinson's disease) conducted between June 2009 and December 2011~\cite{29_yarnall2014characterizing}. This study was conducted according to the declaration of Helsinki and had ethical approval from the Newcastle and North Tyneside research ethics committee (REC reference: 09/H0906/82). All participants signed an informed consent form prior to testing.
%
%
Each participant was asked to wear three Opal inertial sensors (APDM, Inc., Portland, OR, USA): one on the fifth lumbar vertebrae (L5) and two on the ankles. The Opal sensors include a triaxial accelerometer, a gyroscope and a magnetometer and record signal data at \SI{128}{\Hz}.
Traditional gait assessment was concurrently conducted as part of the \mbox{ICICLE-GAIT} study using a \SI{7.0}{\meter} long and \SI{0.6}{\meter} wide instrumented walkway (Platinum model GAITRite, software version 4.5, CIR systems, NJ, USA). The instrumented walkway had a spatial accuracy of \SI{1.27}{\cm} and a temporal accuracy of about \SI{4.17}{\ms} (with a sampling frequency of \SI{240}{\Hz}), 
and was synchronised with the Opal sensors.  
%
Participants were asked to walk at their preferred speed under two conditions: (i) performing four intermittent straight line walking trials (IW) over \SI{10}{\meter} (the instrumented walkway was placed at the centre of the \SI{10}{\meter}~\cite{19_del2016validation}), and (ii) continuously for 2 minutes on a \SI{25}{\meter} oval circuit (CW), where gait was measured only as they walked on the instrumented walkway placed in the middle of the circuit to ensure gait was captured at a steady speed~\cite{30_galna2013gait}.
PD participants who were on medication were tested approximately one hour after medication intake.

\subsection{Data pre-processing}

The Opal sensors, along with the raw inertial data, also provide \mbox{pre-computed} orientation matrices. This information is used in the first part of the data processing phase to project the data onto a reference system that is fixed to the subject, composed by the craniocaudal (longitudinal), anteroposterior (sagittal) and mediolateral (frontal) axes. This process is key to have a consistent reference system among all the measures, and also to compensate for reasonable variations in the sensor placement.
Then, each signal is normalised to have zero mean and unit average power. In this work, both triaxial accelerometer and gyroscope data are considered, including their magnitudes. Considering the three sensor locations, a single sample is composed of $24$ time series ($4$ accelerometer data, accounting for $x$, $y$, $z$ components and the signal's magnitude, and $4$ gyroscope data for each sensor location), which are processed by the model described in the following.


\subsection{The proposed network architecture}
\begin{figure}[tpb]
    \centering
    \includegraphics[width=0.9\columnwidth]{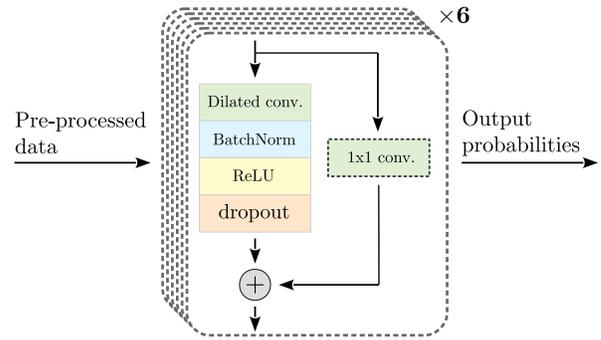}
    \caption{Schematic diagram of the GSN architecture.}
    \label{fig:network_scheme}
\end{figure}

A deep CNN, referred to as gait segmentation network (GSN), has been developed and trained with the purpose of improving traditional gait segmentation from inertial signals. A schematic diagram of the proposed network architecture is shown in \fig{fig:network_scheme}. The network is structured into $6$ consecutive layers. First, the network receives an \mbox{$N$-dimensional} input and it applies a convolution operation with $C$ channels (\mbox{$C=128$} has been considered for the presented results). Then, a stack of five dilated convolutions, with $C$ channels each, follows. Incidentally, dilated convolutions have been proved to be successful for the semantic segmentation of images~\cite{wang2018understanding}. Therefore, here, we thought of extending their capabilities for the detection of events in \mbox{multi-dimensional} inertial signals.

\begin{figure}[tpb]
    \centering
    \includegraphics[width=\columnwidth]{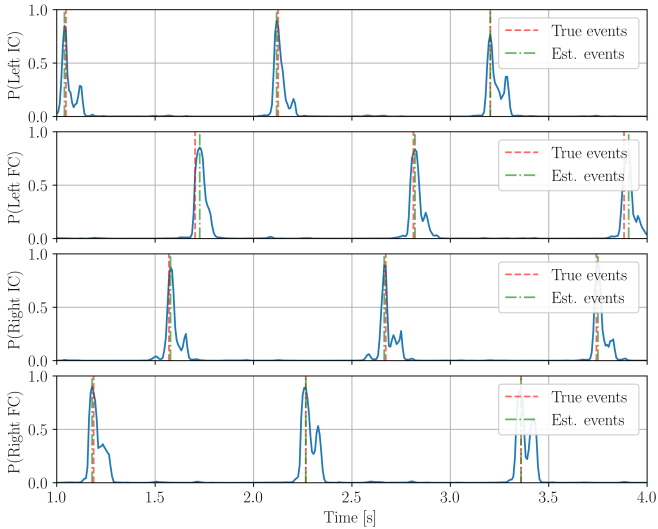}
    \caption{A representative example of gait event identification using the proposed GSN network (blue solid line) and the traditional gold standard method (red dashed line).} 
    \label{fig:output_example}
\end{figure}

The GSN structure is designed so that the output dimensionality of each layer perfectly matches that of its input, without performing data dimensionality reduction or expansion. The dilation coefficient is doubled for each subsequent convolution, so that the receptive field increases exponentially with the number of layers. Each layer implements a residual connection at the output, which allows for a faster and improved convergence~\cite{he2016deep}. Where appropriate, a convolution with a $1\times1$ kernel is applied to the residual branch for dimensionality matching. Batch normalisation is used before each rectified linear unit (ReLU) activation function to increase the training stability, and also for improved regularisation~\cite{ioffe2015batch}.
Four different outputs have been considered for this study: two for the estimation of the right foot initial contact (IC) and final contact (FC), and two for the left foot IC and FC.
A single output estimates the likelihood of the corresponding input sample, given a specific gait event (e.g., the right foot IC). 
\fig{fig:output_example} shows the output corresponding to each of the four considered gait events, with the dashed red vertical lines representing the events detected by the GAITRite standard reference. 
From visual inspection, it can be noted that a higher probability (close to one) is output by the network in the (time) neighbourhood of the correct target event (as detected by the GAITRite)
Incidentally, the exact identification of the events entails the use of an additional peak detection algorithm. Indeed, the events correspond to the likelihood peaks (shown as green dashed vertical lines in \fig{fig:output_example}), with some constraints imposed to avoid multiple peaks corresponding to the same event, and to make the number of false positives and negatives as small as possible.
It is worth noting that the newly proposed architecture is capable of processing variable length input data, with no restriction on the input size. Moreover, a single forward pass allows to estimate all the events, which makes the GSN model quite convenient to use in practical applications. 
Finally, an adaptive moment optimizer has been used to minimize a binary cross-entropy loss function~\cite{kingma2014adam}.

\section{RESULTS}
\label{sec:results}

The portion of data obtained by the GAITRite (both IW and CW types) has been considered for the optimisation process and for the evaluation of the results. A \mbox{$5$-fold} cross validation has been used to increase the reliability of the results. The samples have been grouped so that the data belonging to the same subject were either in the training or in the validation set.
The results are arranged in two sections: the performance in event identification for gait segmentation is analysed in Sec.~\ref{sec:gait_segmentation}, while the capability of GSN to extract \mbox{high-level} gait features is discussed in Sec.~\ref{sec:gait_features}.

\subsection{Gait event identification} 
\label{sec:gait_segmentation}
\begin{figure}[tpb]
    \centering
    \includegraphics[width=\columnwidth]{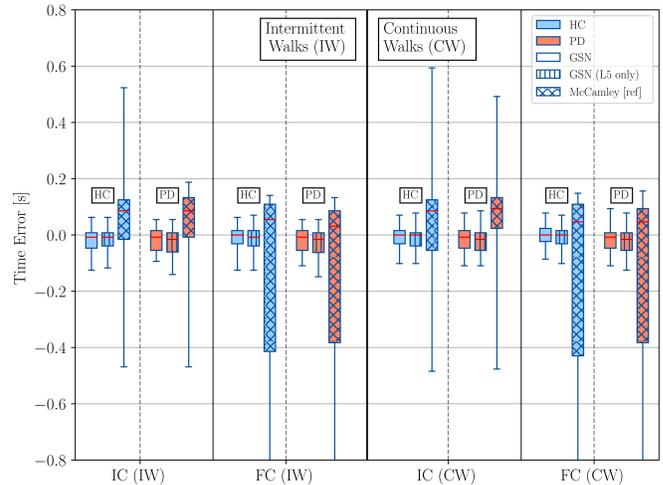}
    \caption{Statistical description of the time error for each gait event, initial (IC) and final contacts (FC), both for HC and PD patients (left and right foot events have been aggregated).
	} 
    \label{fig:time_errors}
\end{figure}

The first result concerns the gait segmentation performance, i.e., the capability of the proposed method to correctly identify gait events (IC and FC for both feet). To quantify the results we compared the estimated events with those localized by the GAITRite reference system. Time errors, i.e., the temporal distance between the estimated and the GAITRite events, are shown in \fig{fig:time_errors}, where the boxes represent the \mbox{inter-quartile} range (IQR), the red lines are the median values, and the whiskers show the $5$th and the $95$th percentile.
\fig{fig:time_errors} is intended to show the clear difference in the variability of results obtained with the gold standard reference and the proposed method, as well as the very good match between outcomes from using a single sensor (on the back).
As a term of comparison, we also implemented one of the most popular approaches for gait segmentation, proposed by McCamley et al., based on WT~\cite{mccamley2012enhanced}.
All the GSN errors are considerably smaller than the WT based method, both in terms of bias (by definition the median value of the error) and IQR, especially for the FC events, for which the WT approach often provides unreliable estimates (e.g., the IQR of the WT based method for FC and IW trials is about \SI{0.52}{\s}, compared to \SI{0.04}{\s} of GSN).
As expected, GSN exhibits slightly worse performance for the PD group, but still with an IQR lower than \SI{70}{\ms} and a bias lower than \SI{8}{\ms} (absolute value).
No statistically significant difference can be observed between IW and CW for all the methods, which are, then, jointly considered in the following results.
Even though the simultaneous analysis of multiple sensors is one of the main benefits of GSN, we also implemented a version of GSN that was solely based on the L5 sensor (termed ``GSN (L5 only)''), in order to provide a fair comparison with the WT based method that solely uses that sensor. The results show that GSN can achieve an almost optimal performance even with a reduced source of data, extending the effectiveness of GSN to a single sensor scenario, that is usually employed in \mbox{free-living} acquisitions. These outcomes support the reliability of GSN as a segmentation and \mbox{gait-related} events identification method, that typically represent the preliminary (and critical) steps for further gait analysis.

\subsection{High-level gait-related features}
\label{sec:gait_features}
Given the above reliable segmentation and event detection performance, next, \mbox{high-level} gait features, such as stride, stance and swing, are estimated and compared with values computed using the GAITRite~\cite{19_del2016validation}.
To this end, we consider the most common \mbox{gait-related} metrics, i.e., the average, variability and asymmetry values, obtained during an entire walking session~\cite{19_del2016validation}. Results are summarized in Tab.~\ref{tab:result_table_time_errors_compact}. Estimation errors are evaluated with respect to the same metrics measured by the gold standard GAITRite reference, and reported in terms of bias and IQR.
A great benefit of GSN is the consistency of the results, which exhibit a very limited number of outliers. Also, negligible performance variations (in terms of estimation errors) can be observed between the HC and PD groups, showing a high reliability even with a challenging dataset which includes pathological individuals. A slight underestimation of the stance and swing variability and asymmetry can be observed (negative bias). 
Concerning the statistical distribution of the features, a small decrease is observed for the IW group with respect to CW, during which the subjects were more likely to achieve a steady state walking. 
Higher values (lower rhythm, higher variability and asymmetry) are also measured for the PD group compared to HC.
An assessment of the statistical significance of these differences provides an interesting avenue to future studies.

\begin{table}[]
	\footnotesize
	\begin{center}
		\setlength\extrarowheight{2pt} 
		\begin{tabular}{ccccc}
			\hline\hline
			&    & \textbf{Average} {[}s{]}   & \textbf{Variability} {[}s{]}  & \textbf{Asymmetry}  {[}s{]} \\
			&    & bias (IQR) & bias (IQR)    & bias (IQR) \\ \hline
			\multirow{2}{*}{\textbf{Stride}} & HC & 0.000 (0.004)      & -0.005 (0.014)       & 0.000 (0.008)      \\ \cline{2-5} 
			& PD & 0.000 (0.008)      & -0.003 (0.016)       & 0.000 (0.012)      \\ \hline
			\multirow{2}{*}{\textbf{Stance}} & HC & -0.008 (0.020)     & -0.012 (0.029)       & -0.008 (0.027)     \\ \cline{2-5} 
			& PD & 0.000 (0.023)      & -0.008 (0.023)       & -0.008 (0.032)     \\ \hline
			\multirow{2}{*}{\textbf{Swing}}  & HC & 0.000 (0.020)      & -0.008 (0.027)       & -0.008 (0.023)     \\ \cline{2-5} 
			& PD & 0.000 (0.020)      & -0.010 (0.023)       & -0.008 (0.031)     \\ \hline\hline
		\end{tabular}
	\end{center}
	\caption{Estimation errors (for IW and CW) with respect to the GAITRite standard reference values of stride, stance and swing (rows), average, variability and asymmetry (columns) metrics. The median value (bias) and the interquartile rage (IQR) are shown. The HC and PD groups are considered separately.}
	\label{tab:result_table_time_errors_compact}
\end{table}


%

\section{CONCLUSION}
\label{sec:conclusions}

In this study, we propose a novel deep learning based method for digital gait segmentation. A comparison with one of the most used algorithms based on wavelet transforms shows the higher reliability of the proposed approach. A rich and challenging dataset, including both healthy controls (HC) and patients suffering from Parkinson's disease (PD) with different severity of motor symptoms, has been used for validation. The performance has been assessed in terms of event identification capabilities (i.e., the effectiveness in the correct time localization of the initial and final foot contact) and reliability in the quantification of the most common \mbox{high-level} \mbox{gait-related} features (average, variability and asymmetry of stride, stance and swing time). The outcomes strongly encourage further studies. Notably, the proposed gait feature extraction model is particularly suitable for use in unsupervised conditions or when a reference system (e.g., the GAITRite) is unavailable. The inclusion of clinical and demographic data of PD and HC groups can be of primary importance for improving the unsupervised assessment of the disease progression, and to assist the clinical personnel to make important therapeutic decisions.

\section{ACKNOWLEDGMENTS}
The ICICLE-GAIT study was supported by Parkinson’s UK (J-0802, G-1301) and by the NIHR Newcastle Biomedical Research Centre. RZUR is supported by the EU H2020 research and innovation program under the Marie Sklodowska-Curie grant agreement No. 721577. SDD is supported by the Newcastle Biomedical Research Centre (BRC) based at Newcastle upon Tyne and Newcastle University. The work was also supported by the NIHR/Wellcome Trust Clinical Research Facility (CRF) infrastructure at Newcastle upon Tyne Hospitals NHS Foundation Trust. All opinions are those of the authors and not the funders. The authors would also like to thank Dr Lisa Alcock and Mr Philip Brown for their assistance with data collection (ICICLE-GAIT study).

{
\linespread{0.988}
\bibliographystyle{IEEEtran}
\bibliography{IEEEabrv,bibliography}
}

\end{document}